\documentstyle[aps,12pt]{revtex}
\begin{document}
\author{J. R. Morris\thanks{%
E-mail : jmorris@iunhaw1.iun.indiana.edu}}
\address{{\it Physics} {\it Department}\\
{\it Indiana University Northwest}\\
{\it 3400 Broadway}\\
{\it Gary, Indiana 46408}
\bigskip\ }
\title{Supersymmetry and Gauge Invariance Constraints in a U(1)$\times $U(1)$%
^{\prime }$-Higgs Superconducting Cosmic String Model}
\maketitle

\begin{abstract}
A supersymmetric extension of the $U(1)\times U(1)^{\prime }$-Higgs bosonic
superconducting cosmic string model is considered, where the action is
constructed from scalar chiral and vector superfields. The chiral
superfields are assumed to transform separately under the Abelian gauge
groups, and the constraints imposed upon such a model due to
renormalizability, supersymmetry, and gauge invariance are examined. For a
simple model with a single $U(1)$ chiral superfield and a single $%
U(1)^{\prime }$ chiral superfield, the Witten mechanism for bosonic
superconductivity (giving rise to long range gauge fields outside of the
string) does not exist without the introduction of extra chiral
supermultiplets. The simplest model that can accommodate the requisite
interactions for a superconducting cosmic string solution with long range $%
U(1)$ gauge fields requires five chiral supermultiplets. An investigation of
the bosonic sector of this model indicates that a solution is admitted
describing a gauge string surrounded by supersymmetric vacuum. By
considering the effects of the gauge string background upon the remaining
scalar fields of the theory, it is concluded that a parameter range exists
for which the $U(1)$ charged scalar fields condense within the core of the
string, making it a superconducting string that can carry charge and/or
current. It is also found that the neutral scalar field can be associated
with particle bound states that are localized within or near the string core.

\newpage\
\end{abstract}

\section{Introduction}

The intriguing possibility exists that the early universe may have undergone
a series of symmetry breaking phase transitions resulting in the production
of topological defects, such as cosmic strings\cite{nielsen}-\cite{ktbook}.
As pointed out by Witten\cite{witten}, it is possible that cosmic strings
may be superconducting, owing to the presence of a charged field condensate
in the core of the string. Superconducting strings can therefore be endowed
with a charge and/or current generating long range gauge fields outside of
the string. It has also been argued that Grand Unified Theory (GUT) scale
cosmic strings may quite generically exhibit superconducting properties\cite
{peter},\cite{davis}. The additional possibility exists that supersymmetry
was physically realized in the early universe, and that the supersymmetry
was broken at the same time as, or subsequent to, the formation of cosmic
strings. The investigation of basic models and scenarios involving
superconducting cosmic strings within a supersymmetric context therefore
seems relevant.

The basic prototype bosonic superconducting string model\cite{witten} is
based upon a $U(1)\times U(1)^{\prime }$ gauge symmetry. By the Higgs
mechanism, the $U(1)^{\prime }$ gauge symmetry is spontaneously broken by
the vacuum state of the theory, giving rise to the existence of a
topologically stable cosmic string solution. While the $U(1)$ symmetry is
respected by the vacuum, it can get broken within the string core to give
rise to a bosonic condensate. Here, attention is focused upon the
supersymmetric extension of the $U(1)\times U(1)^{\prime }$-Higgs bosonic
superconducting string model. In the nonsupersymmetric version of the model
containing one $U(1)$ complex scalar field and one $U(1)^{\prime }$ complex
scalar field, along with $U(1)$ and $U(1)^{\prime }$ vector gauge fields, a
scalar potential can readily be constructed which allows the Witten
mechanism to operate, giving rise to a bosonic superconducting cosmic string
solution. The scalar potential of this model is subject to the constraints
imposed by renormalizability and gauge invariance. In the supersymmetric
version of the $U(1)\times U(1)^{\prime }$-Higgs model, however, the scalar
potential which describes interactions between various fields is derived
from the superpotential and $D$-type auxiliary terms, and there is an
additional constraint imposed by supersymmetry. The supersymmetric version
of the model can therefore have a different appearance from that of the
nonsupersymmetric version. In fact, what is found is that the supersymmetric
version of the model, with one $U(1)$ chiral superfield and one $%
U(1)^{\prime }$ chiral superfield, along with the Abelian gauge superfields,
does not allow the Witten mechanism to exist, so that no superconducting
string solutions with long range gauge fields are supported by this simple
model. The simplest version of the supersymmetric model that can accommodate
the Witten mechanism apparently requires five chiral superfields, i.e. two $%
U(1)$ chiral superfields, two $U(1)^{\prime }$ chiral superfields, and a
neutral chiral superfield, along with the gauge vector superfields. The
superconducting string solution that is admitted by this model is examined.
By focusing upon the bosonic sector of this model, it is found that the two $%
U(1)^{\prime }$ scalar fields conspire to give rise to a gauge string, while
the two $U(1)$ scalar fields form a condensate within the string core, and
the neutral scalar field can be associated with bound states localized
within or near the string core.

The nonsupersymmetric $U(1)\times U(1)^{\prime }$-Higgs model is briefly
reviewed in section II. In section III a supersymmetric Abelian-Higgs model
with one chiral superfield and a vector gauge field admitting a
Nielsen-Olesen cosmic string solution\cite{nielsen} is presented, where
notation and orientation can be established. A supersymmetric extension of
the $U(1)\times U(1)^{\prime }$-Higgs model forms section IV, where the
relevance of supersymmetry and gauge invariance constraints becomes evident.
Section V involves an examination of the superconducting string solution in
this model, and a summary forms section VI.

\section{The U(1)$\times $U(1)$^{\prime }$ - Higgs Bosonic Superconducting
String Model}

Cosmic gauge strings\cite{nielsen}$-$\cite{ktbook} can emerge as solutions
in an Abelian-Higgs model where the Abelian gauge group undergoes a
spontaneous symmetry breaking. Outside the string core, where the scalar
field becomes nonvanishing, the vector gauge field acquires mass so that
there is no long range gauge field generated by the gauge string
configuration. However, a long range gauge field can be generated if an
additional Abelian gauge group is included\cite{witten}. In a $U(1)\times
U(1)^{\prime }$-Higgs model\cite{vsbook}$,$\cite{witten} containing two
complex scalar fields, with each scalar field transforming separately under
the gauge groups, one Abelian gauge group can be spontaneously broken,
giving rise to cosmic strings, while the other gauge group remains unbroken.
With a suitable potential (and within a particular parameter range), the
charged scalar field associated with the unbroken Abelian symmetry can
stabilize the system through the formation of a scalar condensate within the
core of the string, giving rise to a current density which generates a
nonvanishing long range gauge vector field outside of the string. The
Lagrangian for such a model can be displayed as
\begin{equation}
L=(D^\mu \phi )^{*}(D_\mu \phi )+(D^{\prime \mu }\sigma )^{*}(D_\mu ^{\prime
}\sigma )-\frac 14F^{\mu \nu }F_{\mu \nu }-\frac 14B^{\mu \nu }B_{\mu \nu
}-V,  \label{e1}
\end{equation}

\noindent where $D_\mu ^{\prime }\sigma =(\nabla _\mu +igB_\mu )\sigma $ is
the gauge covariant derivative\footnote{$\nabla _\mu $ is the ordinary
spacetime covariant derivative, for use in curvilinear coordinate systems.}
associated with the spontaneously broken $U(1)^{\prime }$ gauge group, $%
D_\mu \phi =(\nabla _\mu +ieA_\mu )\phi $ is the gauge covariant derivative
associated with the $U(1)$ gauge group, which is unbroken outside of the
string, $\sigma $ is the string scalar field which carries a $U(1)^{\prime }$
charge $Q$ and a $U(1)$ charge $q=0$, and $\phi $ is the scalar field
associated with $U(1)$ which carries $U(1)^{\prime }$ and $U(1)$ charges of $%
Q=0$ and $q$, respectively. The field strength tensors are $B_{\mu \nu
}=\partial _\mu B_\nu -\partial _\nu B_\mu $ and $F_{\mu \nu }=\partial _\mu
A_\nu -\partial _\nu A_\mu $.

The potential can be written in the form\footnote{%
Note that the ``bar'' and ``star'' symbols are used interchangeably to
denote complex conjugation, so that $\bar{\phi}=\phi ^{*}$, for example.}
\begin{equation}
V=\frac 12\lambda _\sigma (\bar{\sigma}\sigma -\eta ^2)^2+f(\bar{\sigma}%
\sigma -\eta ^2)\bar{\phi}\phi +m^2\bar{\phi}\phi +\frac 12\lambda _\phi (%
\bar{\phi}\phi )^2.  \label{e2}
\end{equation}

\noindent The vacuum state is described by $|\sigma |=\eta $, $\phi =0$, and
it is assumed that the parameters $\lambda _\sigma $, $\lambda _\phi $, $f$,
and $m$ are real-valued positive quantities. In the string core $\sigma
\rightarrow 0$, and by the Witten mechanism\cite{witten} there exists a
range of parameters for which the condition
\mbox{$\vert$}
$\phi |\rightarrow |\phi _0|$, where
\begin{equation}  \label{e3}
|\phi _0|^2=\frac{(f\eta ^2-m^2)}{\lambda _\phi },
\end{equation}

\noindent is required for stability of the system, provided that $(f\eta
^2-m^2)>0$. The electromagnetic current density $ej_\mu =ie[\bar{\phi}(D_\mu
\phi )-\phi (D_\mu \phi )^{*}]$ can be nonvanishing in the string core and
can therefore generate long range electromagnetic fields outside the string.
These fields allow long range interactions to exist between strings and
between strings and particles carrying a $U(1)$ charge.

It should also be noted that fermionic superconducting strings can exist\cite
{witten}, where fermionic zero modes, rather than a bosonic condensate,
exist within the string core\cite{vsbook}$,$\cite{witten}, with the fermions
carrying a nonzero $U(1)$ charge. These fermion modes can arise from Yukawa
couplings to scalar fields.

\section{Supersymmetric Extension of the Abelian-Higgs Model}

Let us consider a supersymmetric extension of the Abelian-Higgs model\cite
{fayet1}$,$\cite{sbook} which can give rise to gauge string solutions when
the Abelian gauge group is spontaneously broken. The model consists of a
chiral superfield $\Sigma $ coupled to a vector superfield $B$. (We employ
the Wess-Zumino gauge throughout.) The chiral supermultiplet contains a
complex scalar field $\sigma $, a Weyl 2-spinor $\psi _\alpha $, along with
an auxiliary complex scalar field $F$. The chiral superfield has a
superspace representation\cite{sbook}$,$\cite{wbbook}
\begin{equation}
\Sigma (x,\theta ,\bar{\theta})=\sigma (y)+\sqrt{2}\theta \,\psi (y)+\theta
^2F(y),  \label{e4}
\end{equation}

\noindent where $y^\mu =x^\mu +i\theta \sigma ^\mu \bar{\theta}$. In the
Wess-Zumino gauge, the vector supermultiplet contains a ``photon'' $B_\mu $,
the Weyl spinor ``photino'' fields $\lambda _\alpha $, $\bar{\lambda}_{\dot{%
\alpha}}$, and a $D$-term auxiliary field $D(x)$. The vector superfield $B$
has a superspace representation\cite{sbook}$,$\cite{wbbook}
\begin{equation}
B(x,\theta ,\bar{\theta})=-(\theta \sigma ^\mu \bar{\theta})B_\mu
(x)+i\theta ^2\bar{\theta}\,\bar{\lambda}(x)-i\bar{\theta}^2\theta \,\lambda
(x)+\frac 12\theta ^2\bar{\theta}^2D(x),  \label{e5}
\end{equation}

\noindent where $\theta ^2=\theta \theta \equiv \theta ^\alpha \theta
_\alpha $ with $\alpha =1,2$, $\bar{\theta}^2=\bar{\theta}\bar{\theta}\equiv
\bar{\theta}_{\dot{\alpha}}\bar{\theta}^{\dot{\alpha}}$ with $\dot{\alpha}%
=1,2$, and $\theta \lambda =\theta ^\alpha \lambda _\alpha $, $\bar{\theta}%
\bar{\lambda}=\bar{\theta}_{\dot{\alpha}}\bar{\lambda}^{\dot{\alpha}}$.
[Aside from a metric $g_{\mu \nu }$ with signature $(+,-,-,-)$, we adopt the
conventions (see Appendix) and gamma matrices of ref.\cite{sbook}. ] The
Lagrangian, aside from surface terms, is given by\cite{sbook}$,$\cite{wbbook}
\begin{equation}
\begin{array}{ll}
L & =\frac 14\left[ W^\alpha W_\alpha |_{\theta ^2}+\bar{W}_{\dot{\alpha}}%
\bar{W}^{\dot{\alpha}}|_{\bar{\theta}^2}\right] +\left( \bar{\Sigma}%
e^{gQB}\Sigma \right) |_{\theta ^2\bar{\theta}^2} \\
& +W(\Sigma )|_{\theta ^2}+\bar{W}(\bar{\Sigma})|_{\bar{\theta}^2}+\kappa
D(x)
\end{array}
\,\,,  \label{e6}
\end{equation}

\noindent where\cite{wbbook} $W_\alpha =-\frac 14\bar D^2D_\alpha B$ is the
field strength chiral superfield ($D_\alpha $ is the supersymmetric
covariant derivative), $W$ is the superpotential, $W|_{\theta ^2}$
represents the $\theta ^2$ part of $W$, etc., $\kappa D(x)$ is the
Fayet-Illiopoulos\cite{fayet2} $D$-term, with $\kappa =$constant, and $\bar
\Sigma $ is the complex conjugate of $\Sigma $, etc.

Under an Abelian gauge transformation,
\begin{equation}  \label{e7}
\begin{array}{cc}
\Sigma \rightarrow e^{-i\Lambda Q}\Sigma , & \bar \Sigma \rightarrow \bar
\Sigma e^{i\Lambda Q}, \\
B\rightarrow B+\frac ig(\Lambda -\bar \Lambda ), & W_\alpha \rightarrow
W_\alpha ,
\end{array}
\end{equation}

\noindent where $\Lambda (x,\theta ,\bar \theta )$ is a gauge parameter
chiral superfield, and $Q$ is the Abelian charge. The action associated with
the Lagrangian $L$ is invariant under supersymmetry transformations and the
Abelian gauge transformations in (\ref{e7}).

The model can be extended to accommodate $N$ chiral superfields $\Sigma
_i=(\sigma _i,\psi _i,F_i)$, $i=1,\cdot \cdot \cdot ,N$, with associated
Abelian charges $Q_i$, with the replacement $\Sigma \rightarrow \Sigma _i$, $%
Q\rightarrow Q_i$ and summing over the index $i$ in the expression for the
Lagrangian in (\ref{e6}). Solving for the auxiliary fields then yields\cite
{sbook}
\begin{equation}
\bar{F}_i=-\frac{\partial W}{\partial \sigma _i},\,\,\,\,F_i=-\frac{\partial
\bar{W}}{\partial \bar{\sigma}_i},\,\,\,\,D=-\left[ \kappa +\frac 12g%
\mathop{\displaystyle \sum }
\limits_iQ_i\bar{\sigma}_i\sigma _i\right] .  \label{e8}
\end{equation}

The Lagrangian can be written in terms of the component fields as
\begin{equation}  \label{e9}
L=L_B+L_F+L_Y\,\,,
\end{equation}

\noindent where
\begin{equation}
L_B=(\bar{D}^{i\mu }\bar{\sigma}_i)(D_\mu ^i\sigma _i)-\frac 14B^{\mu \nu
}B_{\mu \nu }-V,  \label{e10}
\end{equation}
\begin{equation}
L_F=-i\psi _i\sigma ^\mu \bar{D}_\mu ^i\bar{\psi}_i-i\lambda \sigma ^\mu
\partial _\mu \bar{\lambda}\,\,,  \label{e11}
\end{equation}
\begin{equation}
L_Y=\frac{ig}{\sqrt{2}}Q_i\left[ \bar{\sigma}_i\psi _i\lambda -\sigma _i\bar{%
\psi}_i\bar{\lambda}\right] -\frac 12\left[ \sum\limits_{k,l=1}^N\left(
\frac{\partial ^2W}{\partial \sigma _k\partial \sigma _l}\right) \psi _k\psi
_l+CC\right] ,  \label{e12}
\end{equation}

\noindent where a sum over $i$ is implied, and where $D_\mu ^i=(\nabla _\mu +%
\frac{ig}2Q_iB_\mu )$, $\bar D_\mu ^i=(\nabla _\mu -\frac{ig}2Q_iB_\mu )$
are the ordinary gauge covariant derivatives. The scalar potential in (\ref
{e10}) is given by
\begin{equation}  \label{e13}
V=\sum_i\bar F_iF_i+\frac 12D^2=\sum_i\left| \frac{\partial W}{\partial
\sigma _i}\right| ^2+\frac 12\left[ \kappa +\frac 12g\sum_iQ_i\bar \sigma
_i\sigma _i\right] ^2.
\end{equation}

\noindent For a renormalizable theory (in four spacetime dimensions) the
most general form of the superpotential is described by
\begin{equation}  \label{e14}
W=a_i\Sigma _i+b_{ij}\Sigma _i\Sigma _j+c_{ijk}\Sigma _i\Sigma _j\Sigma
_k\,\,,
\end{equation}

\noindent where $a_i$, $b_{ij}$, and $c_{ijk}$ are constants and a sum over
the contracted indices is implied. The superpotential $W$ is also
constrained by gauge invariance, i.e. each term in $W$ must have a net
charge of zero, so that $a_i=0$ for $Q_i\neq 0$, $b_{ij}=0$ for $Q_i+Q_j\neq
0$, and $c_{ijk}=0$ for $Q_i+Q_j+Q_k\neq 0$.

In the Fayet model\cite{fayet1}$,$\cite{sbook} with a single chiral
superfield $\Sigma $ ,($i=1$), gauge invariance forces the superpotential $W$
to vanish identically, so that the scalar potential reduces to $V=\frac 12%
D^2=\frac 12(\kappa +\frac 12gQ\bar{\sigma}\sigma )^2$. Upon choosing $%
\kappa =-\frac 12gQ\eta ^2<0$, we have
\begin{equation}
V=\frac 18g^2Q^2(\bar{\sigma}\sigma -\eta ^2)^2\,\,,  \label{e15}
\end{equation}

\noindent and the supersymmetric vacuum state, located by $|\sigma |=\eta $,
spontaneously breaks the Abelian gauge symmetry. In the limit of vanishing
fermion fields, we have $L=L_B$, with the scalar potential given by (\ref
{e15}), which coincides with a broken symmetric Abelian-Higgs model, which
admits as a solution, an ordinary Nielsen-Olesen cosmic Abelian gauge
string. With the inclusion of several chiral superfields $\Sigma _i$, the
theory, in general, supports cosmic strings formed through the breaking of
the Abelian gauge symmetry, along with interactions involving charged
particles which transform under the same Abelian gauge group as gives rise
to the strings. However, no long range forces exist, since the vector field $%
B_\mu $ acquires mass. Therefore, even if bosonic condensates or fermionic
zero modes form in the core of a string, they will not give rise to long
range gauge forces outside of the string.

\section{Supersymmetric Extension of the U(1)$\times $U(1)$^{\prime }$ -
Higgs $\,$Model}

In order to have the possibility of describing superconducting gauge strings
in a supersymmetric theory, where the gauge strings support currents that
give rise to long range gauge fields, let us consider the supersymmetric
extension of the $U(1)\times U(1)^{\prime }$-Higgs model described by the
Lagrangian
\begin{equation}  \label{e16}
\begin{array}{ll}
L= & \frac 14\left[ W_A^\alpha W_{A\alpha }|_{\theta ^2}+\bar W_{A\dot \alpha
}\bar W_A^{\dot \alpha }|_{\bar \theta ^2}\right] +\frac 14\left[ W_B^\alpha
W_{B\alpha }|_{\theta ^2}+\bar W_{B\dot \alpha }\bar W_B^{\dot \alpha }|_{%
\bar \theta ^2}\right] \\
& +(\bar \Sigma _ie^{gQ_iB}\Sigma _i)|_{\theta ^2\bar \theta ^2}+(\bar \Phi
_je^{eq_jA}\Phi _j)|_{\theta ^2\bar \theta ^2} \\
& +\kappa _BD_B(x)+\kappa _AD_A(x)+W(\Sigma ,\Phi )|_{\theta ^2}+\bar W(\bar
\Sigma ,\bar \Phi )|_{\bar \theta ^2}
\end{array}
{}.
\end{equation}

\noindent The chiral matter supermultiplets are described by $\Sigma
_i=(\sigma _i,\psi _i,F_i)$, with $i=1,\cdot \cdot \cdot ,M$, and $\Phi
_j=(\phi _j,\chi _j,G_j)$, with $j=1,\cdot \cdot \cdot ,N$, and the two
vector supermultiplets are described by $A=(A_\mu ,\xi _\alpha ,\bar \xi _{%
\dot \alpha },D_A)$ and $B=(B_\mu ,\lambda _\alpha ,\bar \lambda _{\dot
\alpha },D_B)$, where $F_i$, $G_j$, $D_A$, and $D_B$ are the auxiliary
fields. A Fayet-Illiopoulos term associated with each vector superfield has
been included in (\ref{e16}), and $W_{A\alpha }=-\frac 14\bar D^2D_\alpha A$
is the field strength chiral superfield associated with $A$ and $W_{B\alpha
}=-\frac 14\bar D^2D_\alpha B$ is the corresponding field strength chiral
superfield associated with the vector superfield $B$. The chiral superfield $%
\Phi _j$ carries a $U(1)$ charge of $q_j$ and a zero $U(1)^{\prime }$ charge
($Q_j=0$), while the chiral superfield $\Sigma _i$ carries a $U(1)^{\prime }$
charge of $Q_i$ and a zero $U(1)$ charge ($q_i=0$). The $U(1)$ vector gauge
field is $A_\mu $ and the $U(1)^{\prime }$ vector gauge field is $B_\mu $.
The gauge covariant derivatives for the $U(1)$ and $U(1)^{\prime }$ gauge
groups are
\begin{equation}  \label{e17}
D_\mu ^j=(\nabla _\mu +i\frac e2q_jA_\mu ),\,\,\,\,\,\,\,\,D_\mu ^{\prime
\,i}=(\nabla _\mu +i\frac g2Q_iB_\mu ),
\end{equation}

\noindent respectively. The superfield $\Phi _j$ transforms nontrivially
only under the group $U(1)$, while $\Sigma _i$ transforms nontrivially only
under the group $U(1)^{\prime }$, i.e.
\begin{equation}  \label{e18}
\begin{array}{cc}
\Phi _j\rightarrow e^{-i\Lambda _Aq_j}\Phi _j, & \Sigma _i\rightarrow
e^{-i\Lambda _BQ_i}\Sigma _i, \\
A\rightarrow A+\frac ie(\Lambda _A-\bar \Lambda _A), & B\rightarrow B+\frac i%
g(\Lambda _B-\bar \Lambda _B),
\end{array}
\end{equation}

\noindent where $\Lambda _A$ and $\Lambda _B$ are the gauge parameter chiral
superfields associated with the local $U(1)$ and $U(1)^{\prime }$
transformations, respectively.

In terms of the component fields, let us write the Lagrangian as
\begin{equation}  \label{e19}
L=L_B+L_F+L_Y\,\,,
\end{equation}

\noindent where
\begin{equation}
\begin{array}{ll}
L_B= & (\bar{D}^{j\mu }\bar{\phi}_j)(D_\mu ^j\phi _j)+(\bar{D}^{\prime
\,i\mu }\bar{\sigma}_i)(D_\mu ^{\prime \,i}\sigma _i) \\
& -\frac 14F^{\mu \nu }F_{\mu \nu }-\frac 14B^{\mu \nu }B_{\mu \nu }-V,
\end{array}
\label{e20}
\end{equation}
\begin{equation}
\begin{array}{ll}
L_F= & -i\chi _j\sigma ^\mu \bar{D}_\mu ^j\bar{\chi}_j-i\psi _i\sigma ^\mu
\bar{D}_\mu ^{\prime \,i}\psi _i \\
& -i\xi \sigma ^\mu \nabla _\mu \bar{\xi}-i\lambda \sigma ^\mu \nabla _\mu
\bar{\lambda}\,\,,
\end{array}
\label{e21}
\end{equation}
\begin{equation}
\begin{array}{ll}
L_Y= & \frac{ie}{\sqrt{2}}q_j\left[ \bar{\phi}_j\chi _j\xi -\phi _j\bar{\chi}%
_j\bar{\xi}\right] +\frac{ig}{\sqrt{2}}Q_i\left[ \bar{\sigma}_i\psi
_i\lambda -\sigma _i\bar{\psi}_i\bar{\lambda}\right] \\
& -\frac 12\left[ \sum\limits_{k,l=1}^M\left( \frac{\partial ^2W}{\partial
\sigma _k\partial \sigma _l}\right) \psi _k\psi
_l+\sum\limits_{k,l=1}^N\left( \frac{\partial ^2W}{\partial \phi _k\partial
\phi _l}\right) \chi _k\chi _l+CC\right] ,
\end{array}
\label{e22}
\end{equation}

\noindent with a sum over $i$ and $j$ implied, and where $F_{\mu \nu
}=\partial _\mu A_\nu -\partial _\nu A_\mu $ and $B_{\mu \nu }=\partial _\mu
B_\nu -\partial _\nu B_\mu $, and $CC$ stands for complex conjugate. The
scalar potential is
\begin{equation}
V=\sum_i\bar{F}_iF_i+\sum_j\bar{G}_jG_j+\frac 12D_A^2+\frac 12D_B^2\,\,,
\label{e23}
\end{equation}

\noindent with
\begin{equation}  \label{e24}
\begin{array}{c}
\bar F_i=- \frac{\partial W}{\partial \sigma _i}\,\,,\,\,\,\,\,\,\,\,\bar G%
_j=-\frac{\partial W}{\partial \phi _j}\,\,, \\
D_A=-\left[ \kappa _A+\frac 12e\sum_jq_j\bar \phi _j\phi _j\right]
,\,\,\,\,\,D_B=-\left[ \kappa _B+\frac 12g\sum_iQ_i\bar \sigma _i\sigma
_i\right] .
\end{array}
\end{equation}

\noindent For a renormalizable theory (in four dimensions) the
superpotential $W$, constructed from the chiral superfields \thinspace $%
\Sigma _i$ and $\Phi _j$, has the general form
\begin{equation}  \label{e25}
W=a_iX_i+b_{ij}X_iX_j+c_{ijk}X_iX_jX_k\,\,,
\end{equation}

\noindent where $X_i$ represents either $\Sigma _i$ or $\Phi _i$, and $W$ is
again constrained by gauge invariance, so that each term must have a total $%
U(1)$ charge of zero ($\sum q_i=0$) and a total $U(1)^{\prime }$ charge of
zero ($\sum Q_i=0$).

For the simplest supersymmetric analog of the model described by (\ref{e1}),
corresponding to the case where there is a single chiral superfield $\Sigma $
along with a single chiral superfield $\Phi $, we can choose $\kappa _A\ge 0$
and $\kappa _B=-\frac 12gQ\eta ^2<0$, so that the $U(1)$ symmetry is
unbroken and the $U(1)^{\prime }$ symmetry is spontaneously broken. We then
have a model containing cosmic gauge strings along with $U(1)$ charged
particles. However, the only interactions between the supermultiplets $%
\Sigma $ and $\Phi $ must come from the superpotential $W$, given by (\ref
{e25}). But, by gauge invariance, $W=0$, so that no interactions exist
between $U(1)$ charged particles and a $U(1)^{\prime }$ gauge string. The
Witten mechanism, therefore, does not exist in this simple model. Also, note
that if $|\sigma |=\eta $ in the vacuum state, and $\kappa _A=0$, then $V=0$
in the vacuum state and the supersymmetry is not broken in the vacuum.
However, for the case where $|\sigma |=\eta $ in the vacuum state and $%
\kappa _A>0$, then $V>0$ in the vacuum state and supersymmetry is
spontaneously broken.

Evidently, from (\ref{e19}) - (\ref{e25}), the only way to generate an
interaction of $U(1)$ $\Phi $ fields with $U(1)^{\prime }$ $\Sigma $ fields,
which can result in a theory of gauge strings possessing $U(1)$ charge
and/or current, is to build an interaction between the $\Sigma $ and $\Phi $
fields through a nonvanishing superpotential. The form of the
superpotential, dictated by the initial requirement of a supersymmetric
action, along with the constraint imposed by gauge invariance, apparently
requires a more complicated supersymmetric version of the model as compared
to the nonsupersymmetric version. For example, in the supersymmetric $%
U(1)\times U(1)^{\prime }$-Higgs model above, it would appear that the
simplest model allowing interactions between $\Phi $ fields and $\Sigma $
fields, due to interaction terms in the superpotential $W$, would consist of
five chiral superfields, of which two have nonzero $U(1)^{\prime }$ charges,
two have nonzero $U(1)$ charges, and one neutral chiral superfield ${\cal Z}%
=(Z,\psi _Z,F_Z)$ has zero $U(1)$ charge and zero $U(1)^{\prime }$ charge ($%
q_Z=Q_Z=0$). Such a set of fields allows the superpotential terms
\begin{equation}
W_{int}=c_1{\cal Z}\Sigma _1\Sigma _2+c_2{\cal Z}\Phi _1\Phi _2\,\,,
\label{e26}
\end{equation}

\noindent with $Q_{\Sigma _1}=-Q_{\Sigma _2}$, $q_{\Phi _1}=-q_{\Phi _2}$,
giving rise to the scalar potential terms
\begin{equation}  \label{e27}
\begin{array}{ll}
V_{int} & =\left| c_1\sigma _1\sigma _2+c_2\phi _1\phi _2\right| ^2 \\
& +\left| Z\right| ^2\left[ c_1^2\left( \left| \sigma _1\right| ^2+\left|
\sigma _2\right| ^2\right) +c_2^2\left( \left| \phi _1\right| ^2+\left| \phi
_2\right| ^2\right) \right] ,
\end{array}
\end{equation}

\noindent along with Yukawa terms of the form
\begin{equation}  \label{e28}
L_{Y,\,int}=-\frac 12\left( \frac{\partial ^2W}{\partial X_i\partial X_j}%
\right) \Psi _i\Psi _j+CC,
\end{equation}

\noindent where a sum over $i$ and $j$ is implied, $CC$ stands for complex
conjugate, $X_i$ represents $\sigma _i$, $\phi _i$, or $Z$, and $\Psi _i$
represents $\psi _i$, $\chi _i$, or $\psi _Z$ (the spinor component of the $%
Z $ chiral supermultiplet). Pure $Z$ terms, such as $a_Z{\cal Z}+b_Z{\cal Z}%
^2+c_Z{\cal Z}^3$, can also be included in $W$, resulting in additional
contributions to the scalar potential $V$ and additional Yukawa terms in $%
L_Y $. The possibility may then exist for destabilizing the set of states
described by $\phi _j=0$ in the core of a gauge string, thereby allowing a $%
U(1)$ charged bosonic condensation and/or the formation of $U(1)$ charged
fermionic zero modes in the string core.

\section{A Superconducting Cosmic String Solution in the Supersymmetric U(1)$%
\times $U(1)$^{\prime }$-Higgs Model}

Let us now consider the type of model discussed in section IV, where five
chiral superfields are introduced so that a nonvanishing superpotential can
be constructed which allows interactions between the $U(1)^{\prime }$ fields
and the $U(1)$ fields. Specifically, we consider a set of fields such that
two of them ($\Sigma _i$) have nonzero $U(1)^{\prime }$ charges and zero $%
U(1)$ charges, two other fields ($\Phi _i$) have nonzero $U(1)$ charges and
zero $U(1)^{\prime }$ charges, and one field (${\cal Z}$) has a zero $%
U(1)^{\prime }$ charge and a zero $U(1)$ charge. The Fayet-Illiopoulos terms
will be dropped, and it can be seen that there exists a supersymmetric
vacuum state for the theory in which the $U(1)^{\prime }$ symmetry is
spontaneously broken and the $U(1)$ symmetry is unbroken. The spontaneous
breaking of the $U(1)^{\prime }$ symmetry indicates the existence of a
topologically stable gauge string. The gauge string is described by the
bosonic components of the $\Sigma _i$ fields (denoted by $\sigma _i$) along
with the $U(1)^{\prime }$ gauge field $B_\mu $. By focusing on the bosonic
sector of the theory, where the fermion fields vanish, the stability of the
bosonic components of the $\Phi _i$ and ${\cal Z}$ superfields (denoted by $%
\phi _i$ and $Z$, respectively) in the presence of the gauge string
background can be examined. Following the type of argument used by Witten%
\cite{witten}, it is argued that a parameter range can exist such that, at
least within a first approximation, the $U(1)$ fields $\phi _i$ form a
condensate within the string core, while the neutral field $Z$ can be
associated with bound states localized within or near the core of the string.

\subsection{Fields, Charge Assignments, and Superpotential}

Let us denote the $U(1)^{\prime }$ charge of a field by $Q_i$ and the $U(1)$
charge of a field by $q_i$. Consider now a model of the type described in
section IV characterized by the five chiral superfields $\Sigma _{\pm }$, $%
\Phi _{\pm }$, and ${\cal Z}$. The charges for the $\Sigma _{\pm }$ fields
are taken to be $(Q_{\pm },q_{\pm })=(\pm 1,0)$, those for the $\Phi _{\pm }$
fields are taken to be $(Q_{\pm },q_{\pm })=(0,\pm 1)$, and for the neutral $%
{\cal Z}$ field we have $(Q_Z,q_Z)=(0,0)$. The superfields can be displayed
in terms of their component fields:
\begin{equation}
\begin{array}{c}
\Sigma _{\pm }=(\sigma _{\pm },\psi _{\pm },F_{\sigma \pm }),\,\,\,\,\,\Phi
_{\pm }=(\phi _{\pm },\chi _{\pm },F_{\phi \pm }),\,\,\,\,\,{\cal Z}=(Z,\psi
_Z,F_Z), \\
A=(A_\mu ,\xi _\alpha ,\bar{\xi}_{\dot{\alpha}},D_A),\,\,\,\,\,B=(B_\mu
,\lambda _\alpha ,\bar{\lambda}_{\dot{\alpha}},D_B).
\end{array}
\label{eq1}
\end{equation}

\noindent We assume that the Fayet-Illiopolous terms vanish, i.e. $\kappa
_A=\kappa _B=0$. The superpotential for the model is assumed to be given by
\begin{equation}
W=\lambda {\cal Z}(\Sigma _{+}\Sigma _{-}-\eta ^2)+c{\cal Z}\Phi _{+}\Phi
_{-}+m\Phi _{+}\Phi _{-}\,\,,  \label{eq2}
\end{equation}

\noindent and the parameters $\lambda $, $c$, $m$, and $\eta $ are assumed
to be real-valued positive quantities.

The scalar potential $V$ is then given by
\begin{equation}
V=\sum_i|F_i|^2+\frac 12D_A^2+\frac 12D_B^2,  \label{eq3}
\end{equation}

\noindent where
\begin{equation}
D_A=-\frac 12e(\bar{\phi}_{+}\phi _{+}-\bar{\phi}_{-}\phi
_{-}),\,\,\,\,\,\,\,\,\,\,D_B=-\frac 12g(\bar{\sigma}_{+}\sigma _{+}-\bar{%
\sigma}_{-}\sigma _{-}),  \label{eq4}
\end{equation}
\begin{equation}
\bar{F}_i=-\frac{\partial W}{\partial X_i},\,\,\,\,\,\,\,\,\,\,X_i=\sigma
_i,\phi _i,{\cal Z}.  \label{eq5}
\end{equation}

\noindent The scalar potential can therefore be written fully as
\begin{equation}
\begin{array}{ll}
V= & \lambda ^2(\bar{\sigma}_{+}\bar{\sigma}_{-}-\eta ^2)(\sigma _{+}\sigma
_{-}-\eta ^2)+\lambda c(\bar{\sigma}_{+}\bar{\sigma}_{-}-\eta ^2)\phi
_{+}\phi _{-}+\lambda c(\sigma _{+}\sigma _{-}-\eta ^2)\bar{\phi}_{+}\bar{%
\phi}_{-} \\
& +c^2\bar{\phi}_{+}\phi _{+}\bar{\phi}_{-}\phi _{-}+\lambda ^2\bar{Z}Z(\bar{%
\sigma}_{+}\sigma _{+}+\bar{\sigma}_{-}\sigma _{-})+(c\bar{Z}+m)(cZ+m)(\bar{%
\phi}_{+}\phi _{+}+\bar{\phi}_{-}\phi _{-}) \\
& +\frac{e^2}8(\bar{\phi}_{+}\phi _{+}-\bar{\phi}_{-}\phi _{-})^2+\frac{g^2}8%
(\bar{\sigma}_{+}\sigma _{+}-\bar{\sigma}_{-}\sigma _{-})^2.
\end{array}
\label{eq6}
\end{equation}

{}From (\ref{eq3}) it can be seen that the lowest energy state of the theory
is a supersymmetric vacuum state with $V=0$, determined by the conditions $%
F_i=0$ and $D_A=D_B=0$. Specifically, the supersymmetric vacuum state that
spontaneously breaks the $U(1)^{\prime }$ symmetry but respects the $U(1)$
symmetry is characterized by
\begin{equation}
|\sigma _{\pm }|_0=\eta ,\,\,\,\,\,\,\,\,\,(\phi _{\pm
})_0=0,\,\,\,\,\,\,\,\,\,Z_0=0.  \label{eq7}
\end{equation}

\subsection{The Bosonic Sector}

\subsubsection{Lagrangian}

Let us now focus upon the bosonic sector of the theory, where the fermion
fields are assumed to vanish. The Lagrangian for the bosonic fields can be
displayed as
\begin{equation}
\begin{array}{ll}
L_B= & (\bar{D}^{\prime \,+\mu }\bar{\sigma}_{+})(D^{\prime }\,_\mu
^{+}\sigma _{+})+(\bar{D}^{\prime \,-\mu }\bar{\sigma}_{-})(D_\mu ^{\prime
\,-}\sigma _{-}) \\
& +(\bar{D}^{+\mu }\bar{\phi}_{+})(D_\mu ^{+}\phi _{+})+(\bar{D}^{-\mu }\bar{%
\phi}_{-})(D_\mu ^{-}\phi _{-})+\partial ^\mu \bar{Z}\partial _\mu Z \\
& -V-\frac 14F^{\mu \nu }F_{\mu \nu }-\frac 14B^{\mu \nu }B_{\mu \nu }\,\,,
\end{array}
\label{eq8}
\end{equation}

\noindent where
\begin{equation}
\begin{array}{ll}
(D_\mu ^{\prime \,\pm }\sigma _{\pm })=(\nabla _\mu \pm \frac{ig}2B_\mu
)\sigma _{\pm }\,\,, & (\bar{D}_\mu ^{\prime \,\pm }\bar{\sigma}_{\pm
})=(\nabla _\mu \mp \frac{ig}2B_\mu )\bar{\sigma}_{\pm }=(D_\mu ^{\prime
\,\pm }\sigma _{\pm })^{*} \\
(D_\mu ^{\pm }\phi _{\pm })=(\nabla _\mu \pm \frac{ie}2A_\mu )\phi _{\pm
}\,\,, & \,(\bar{D}_\mu ^{\pm }\bar{\phi}_{\pm })=(\nabla _\mu \mp \frac{ie}2%
A_\mu )\bar{\phi}_{\pm }=(D_\mu ^{\pm }\phi _{\pm })^{*}\,\,.
\end{array}
\label{eq9}
\end{equation}

\noindent (Note that, due to the convention in defining the gauge
transformations [see (\ref{e18})], the ``physical'' charges are $\tilde{g}%
=g/2$ and $\tilde{e}=e/2$.)

\subsubsection{Field Equations}

{}From the Lagrangian given by (\ref{eq8}) the equations of motion for the
fields $\sigma _{+}$, $\sigma _{-}$, $\phi _{+}$, $\phi _{-}$, $Z$, $A_\mu $%
, and $B_\mu $ can be obtained:
\begin{equation}
\begin{array}{l}
(\nabla _\mu +\frac{ig}2B_\mu )(\nabla ^\mu +\frac{ig}2B^\mu )\sigma
_{+}+\lambda ^2\bar{\sigma}_{-}(\sigma _{+}\sigma _{-}-\eta ^2) \\
+\lambda c\bar{\sigma}_{-}\phi _{+}\phi _{-}+\lambda ^2\bar{Z}Z\sigma _{+}+%
\frac{g^2}4\sigma _{-}(\bar{\sigma}_{+}\sigma _{+}-\bar{\sigma}_{-}\sigma
_{-})=0\,\,,
\end{array}
\label{eq10}
\end{equation}
\begin{equation}
\begin{array}{l}
(\nabla _\mu -\frac{ig}2B_\mu )(\nabla ^\mu -\frac{ig}2B^\mu )\sigma
_{-}+\lambda ^2\bar{\sigma}_{+}(\sigma _{+}\sigma _{-}-\eta ^2) \\
+\lambda c\bar{\sigma}_{+}\phi _{+}\phi _{-}+\lambda ^2\bar{Z}Z\sigma _{-}-%
\frac{g^2}4\sigma _{-}(\bar{\sigma}_{+}\sigma _{+}-\bar{\sigma}_{-}\sigma
_{-})=0\,\,,
\end{array}
\label{eq11}
\end{equation}
\begin{equation}
\begin{array}{l}
(\nabla _\mu +\frac{ie}2A_\mu )(\nabla ^\mu +\frac{ie}2A^\mu )\phi
_{+}+\lambda c\bar{\phi}_{-}(\sigma _{+}\sigma _{-}-\eta ^2)+c^2\bar{\phi}%
_{-}\phi _{+}\phi _{-} \\
+(c\bar{Z}+m)(cZ+m)\phi _{+}+\frac{e^2}4\phi _{+}(\bar{\phi}_{+}\phi _{+}-%
\bar{\phi}_{-}\phi _{-})=0\,\,,
\end{array}
\label{eq12}
\end{equation}
\begin{equation}
\begin{array}{l}
(\nabla _\mu -\frac{ie}2A_\mu )(\nabla ^\mu -\frac{ie}2A^\mu )\phi
_{-}+\lambda c\bar{\phi}_{+}(\sigma _{+}\sigma _{-}-\eta ^2)+c^2\bar{\phi}%
_{+}\phi _{+}\phi _{-} \\
+(c\bar{Z}+m)(cZ+m)\phi _{-}-\frac{e^2}4\phi _{-}(\bar{\phi}_{+}\phi _{+}-%
\bar{\phi}_{-}\phi _{-})=0\,\,,
\end{array}
\label{eq13}
\end{equation}
\begin{equation}
\nabla _\mu \nabla ^\mu Z+\lambda ^2Z(\bar{\sigma}_{+}\sigma _{+}+\bar{\sigma%
}_{-}\sigma _{-})+c(cZ+m)(\bar{\phi}_{+}\phi _{+}+\bar{\phi}_{-}\phi
_{-})=0\,\,,  \label{eq14}
\end{equation}
\begin{equation}
\begin{array}{ll}
\Box B_\mu & =J_{B\mu }\,\, \\
& =\frac{ig}2\left\{ [\bar{\sigma}_{+}(D_\mu ^{\prime \,+}\sigma
_{+})-\sigma _{+}(\bar{D}_\mu ^{\prime \,+}\bar{\sigma}_{+})]-[\bar{\sigma}%
_{-}(D_\mu ^{\prime \,-}\sigma _{-})-\sigma _{-}(\bar{D}_\mu ^{\prime \,-}%
\bar{\sigma}_{-})]\right\} \,\,,
\end{array}
\label{eq15}
\end{equation}
\begin{equation}
\begin{array}{ll}
\Box A_\mu & =J_{A\mu } \\
& =\frac{ie}2\left\{ [\bar{\phi}_{+}(D_\mu ^{+}\phi _{+})-\phi _{+}(\bar{D}%
_\mu ^{+}\bar{\phi}_{+})]-[\bar{\phi}_{-}(D_\mu ^{-}\phi _{-})-\phi _{-}(%
\bar{D}_\mu ^{-}\bar{\phi}_{-})]\right\} \,\,,
\end{array}
\label{eq16}
\end{equation}

\noindent where $\Box =\nabla _\mu \nabla ^\mu =(\partial _0^2-\nabla ^2)$,
and the Lorentz gauges $\nabla _\mu B^\mu =0$, $\nabla _\mu A^\mu =0$ have
been used.

\subsubsection{Simplifying Ansatz}

In order to simplify the system considered above, let us implement a
particular ansatz for the scalar fields for which
\begin{equation}
\phi _{+}=\phi ,\,\,\,\,\,\phi _{-}=\bar{\phi},\,\,\,\,\,\sigma _{+}=\sigma
,\,\,\,\,\,\sigma _{-}=\bar{\sigma},  \label{eq17}
\end{equation}

\noindent i.e. $\phi _{+}$ and $\phi _{-}$ are complex conjugates of one
another, as are $\sigma _{+}$ and $\sigma _{-}$, so that the set of five
complex scalar fields $\{\sigma _{+},\sigma _{-},\phi _{+},\phi _{-},Z\}$
reduces to the set of three complex scalar fields $\{\sigma ,\phi ,Z\}$. The
field equations (\ref{eq10})-(\ref{eq16}) then reduce to
\begin{equation}
(\nabla _\mu +\frac{ig}2B_\mu )(\nabla ^\mu +\frac{ig}2B^\mu )\sigma +\sigma
\left[ \lambda ^2(\bar{\sigma}\sigma -\eta ^2)+\lambda c\bar{\phi}\phi
+\lambda ^2\bar{Z}Z\right] =0,  \label{eq18}
\end{equation}
\begin{equation}
(\nabla _\mu +\frac{ie}2A_\mu )(\nabla ^\mu +\frac{ie}2A^\mu )\phi +\phi
\left[ \lambda c(\bar{\sigma}\sigma -\eta ^2)+c^2\bar{\phi}\phi +(c\bar{Z}%
+m)(cZ+m)\right] =0,  \label{eq19}
\end{equation}
\begin{equation}
\nabla _\mu \nabla ^\mu Z+2\lambda ^2Z\bar{\sigma}\sigma +2c(cZ+m)\bar{\phi}%
\phi =0,  \label{eq20}
\end{equation}
\begin{equation}
\Box B_\mu =J_{B\mu }=ig\left[ \bar{\sigma}(D_\mu ^{\prime }\sigma )-\sigma
(D_\mu ^{\prime }\sigma )^{*}\right] ,  \label{eq21}
\end{equation}
\begin{equation}
\Box A_\mu =J_{A\mu }=ie\left[ \bar{\phi}(D_\mu \phi )-\phi (D_\mu \phi
)^{*}\right] ,  \label{eq22}
\end{equation}

\noindent where $D_\mu ^{\prime }=(\nabla _\mu +\frac{ig}2B_\mu )$ and $%
D_\mu =(\nabla _\mu +\frac{ie}2A_\mu )$.

With the simplifying ansatz given by (\ref{eq17}), the Lagrangian $L_B$
given by (\ref{eq8}) takes the form of an effective Lagrangian given by
\begin{equation}
\begin{array}{ll}
L_{B,eff}= & 2(D^{\prime \,\mu }\sigma )^{*}(D_\mu ^{\prime }\sigma
)+2(D^\mu \phi )^{*}(D_\mu \phi )+\partial ^\mu Z^{*}\partial _\mu Z \\
& -U(\sigma ,\phi ,Z)-\frac 14B^{\mu \nu }B_{\mu \nu }-\frac 14F^{\mu \nu
}F_{\mu \nu }\,\,,
\end{array}
\label{eq23}
\end{equation}

\noindent where
\begin{equation}
\begin{array}{ll}
U(\sigma ,\phi ,Z)= & \lambda ^2(\bar{\sigma}\sigma -\eta ^2)^2+2\lambda c(%
\bar{\sigma}\sigma -\eta ^2)\bar{\phi}\phi +c^2(\bar{\phi}\phi )^2 \\
& +2\lambda ^2\bar{\sigma}\sigma \bar{Z}Z+2\bar{\phi}\phi (c\bar{Z}%
+m)(cZ+m)\,\,.
\end{array}
\label{eq24}
\end{equation}

\noindent It can be verified that the field equations (\ref{eq18})-(\ref
{eq22}) can be obtained directly from $L_{B,eff}$. It can be noted that,
with a rescaling of the fields $\sigma $ and $\phi $, $L_{B,eff\text{ }}$
contains a Lagrangian for the bosonic sector of an ordinary
(nonsupersymmetric) $U(1)\times U(1)^{\prime }$-Higgs model, described in
section II, along with dynamical and interaction terms for the field $Z$.
This observation leads us to strongly suspect that the supersymmetric model
considered here will admit a superconducting cosmic string solution. By
examining the behavior of the fields $\phi $ and $Z$ in the background of
the gauge string formed by the fields $\sigma $ and $B_\mu $, we will see
that, for a suitable parameter range, such a solution does indeed exist,
wherein the field $\phi $ forms a condensate within the string core and the
field $Z$ can be associated with bound states that are localized within or
near the string core.

\subsection{The Superconducting String}

In the vacuum state of the theory we have $|\sigma |=\eta $, $\phi =0$, and $%
Z=0$, so that the $U(1)^{\prime }$ symmetry is spontaneously broken,
allowing a topological gauge string to form. Using cylindrical coordinates $%
(r,\theta ,z)$, let us consider the case in which the fields $\phi $, $Z$,
and $A_\mu $ vanish identically. In this case, we can write $\sigma =\sigma
_s(r,\theta )=s(r)e^{i\theta }$, $B_\mu =\frac 2gB(r)\delta _\mu ^\theta $,
with the boundary conditions
\begin{equation}
\begin{array}{ll}
s(r)\rightarrow \eta ,\,\,\,\,\,B(r)\rightarrow -1, & \text{as\thinspace
\thinspace \thinspace \thinspace \thinspace }r\rightarrow \infty ,. \\
s(r)\rightarrow 0,\,\,\,\,\,B(r)\rightarrow 0, & \text{as}%
\,\,\,\,\,r\rightarrow 0.
\end{array}
\label{eq25}
\end{equation}

\noindent The field equations given by (\ref{eq18}) and (\ref{eq21}),
subject to the boundary conditions in (\ref{eq25}), then admit an ordinary
Nielsen-Olesen Abelian gauge string solution.

Now consider the field $\phi $ in the gauge string background with $Z=0$ and
$\sigma =\sigma _s(r,\theta )$. Following a similar line of reasoning as
that used by Witten\cite{witten}, it can be argued that there exists a
parameter range for which the complex scalar field given by $\phi =0$, with $%
A_\mu =0$, is unstable in the background field of the gauge string, and must
therefore relax to a lower energy state for which $\phi \neq 0$ in the
string core. The first thing that can be noticed is that the potential in
the string core, where $\sigma \rightarrow 0$, is minimized by a value
\begin{equation}
|\phi _0|=\frac{(\lambda c\eta ^2-m^2)^{\frac 12}}c,  \label{eq26}
\end{equation}

\noindent provided that $(\lambda c\eta ^2-m^2)>0$, as will be assumed to be
the case. However, since $\phi \rightarrow 0$ outside of the string core in
the vacuum region, there is also gradient energy to be considered. To see
that the field $\phi $ does assume a nonzero value in the string core, we
set $\sigma =\sigma _s$, $A_\mu =0$, $Z=0$, linearize the equations of
motion, and examine small fluctuations of $\phi $ about $\phi =0$. Writing $%
\phi (\vec{r},t)=\varphi (\vec{r})e^{-i\omega \,t}$, (\ref{eq19}) then gives
\begin{equation}
-\nabla ^2\varphi +\lambda c|\sigma _s|^2\varphi =E\varphi
,\,\,\,\,\,\,\,\,\,\,E\equiv \omega ^2+(\lambda c\eta ^2-m^2),  \label{eq27}
\end{equation}

\noindent which is a Schrodinger-like equation for a particle with ``mass'' $%
\mu =\frac 12$ and ``energy'' $E=\omega ^2+(\lambda c\eta ^2-m^2)$ in a
potential well $\lambda c|\sigma _s|^2$. For the parameter range in which $%
E<(\lambda c\eta ^2-m^2)$, there should exist at least one normalizable
bound state for which $\omega ^2<0$, in which case $\phi =0$ is unstable in
the string core, and therefore the field $\phi $ forms a condensate within
the string, with $\phi \rightarrow \phi _0$ in the core. Furthermore, we
expect, on the basis of continuity, that solutions corresponding to
excitations exist for $A_\mu \neq 0$. These excitations, which can give rise
to charge and current within the string, can be characterized by the
parametrizations $\phi (r,z,t)=F(r)e^{i\psi (z,t)}$, $A_\mu (r,z,t)=\frac 2e%
[P(r)-1]\partial _\mu \psi (z,t)$. For the case in which $\psi =\kappa z\pm
\omega t$, for example, the electromagnetic current density $J_{A\mu }=-%
\frac e2F^2(r)P(r)\partial _\mu \psi $ describes a time independent,
nondissipative charge and current localized within the string core.

At the same level of approximation, let us examine the $Z$ field in the
string background by setting $\sigma =\sigma _s$, $\phi =0$, and $A_\mu =0$,
i.e., as a first approximation we examine the fields $\phi $ and $Z$ in the
absence of one another in the gauge string background. Writing $Z(\vec{r}%
,t)=Z_1(\vec{r})\,e^{-i(\Omega t-kz)}$ , equation (\ref{eq20}) gives
\begin{equation}
-(\partial _x^2+\partial _y^2)Z_1+2\lambda ^2|\sigma _s(r)|^2Z_1=\mu
^2Z_1,\,\,\,\,\,\,\,\,\,\,\mu ^2\equiv \Omega ^2-k^2,  \label{eq28}
\end{equation}

\noindent which is a Schrodinger-like equation for a particle with
``energy'' $\mu ^2$ in the potential well $2\lambda ^2|\sigma _s|^2$. This
attractive potential can accommodate one or more normalizable bound states
with $0<\mu <\sqrt{2}\lambda \eta $, allowing us to infer that $Z$ particles
can be localized within or near the string core in the form of string-$Z$%
-particle bound states. A set of scattering states can evidently exist as
well for $\mu >\sqrt{2}\lambda \eta $.

{}From this level of approximation it is therefore concluded that the $\phi $
condensate gives rise to a superconducting gauge string which may also
accommodate $Z$ particle bound states that can drift along the string.

\section{Summary}

It is possible that supersymmetry was physically realized in the early
universe, and that supersymmetry was broken at the same time as, or
subsequent to, the formation of topological defects, such as cosmic strings.
It therefore seems relevant to investigate basic models and scenarios
involving defects within a supersymmetric context. Here, attention has been
focused upon a relatively simple ``toy'' scenario involving superconducting
Abelian gauge strings with long range $U(1)$ fields. Specifically, a
supersymmetric $U(1)\times U(1)^{\prime }$-Higgs model has been considered,
where the action is constructed from scalar chiral superfields and vector
superfields, with each chiral superfield transforming separately under the
Abelian gauge groups. The $U(1)^{\prime }$ symmetry can be spontaneously
broken, giving rise to gauge strings, while the $U(1)$ symmetry remains
unbroken, giving rise to long range interactions outside of the string. In
the nonsupersymmetric version of the model, a suitable potential can readily
be constructed from two complex scalar fields, which, by the Witten
mechanism, gives rise to a model admitting as solutions bosonic
superconducting strings with long range gauge fields. The basic constraints
imposed upon this potential include those of renormalizability and gauge
invariance. In the supersymmetric extension of this type of model, however,
supersymmetry provides an additional constraint, which is partially
reflected in the form of the superpotential, and therefore in the form of
the scalar potential of the model. For such a model involving only one
chiral superfield $\Phi $, transforming under the $U(1)$ gauge group, and
one chiral superfield $\Sigma $, transforming under the $U(1)^{\prime }$
gauge group, it is seen that no Witten mechanism exists for the formation of
$U(1)$ charged bosonic condensates or fermionic zero modes within a string
which can give rise to long range gauge fields outside of the string.
Apparently, the simplest version of the model that allows interactions
between the $\Sigma $ fields and the $\Phi $ fields involves at least five
chiral superfields. The constraints imposed by supersymmetry and gauge
invariance therefore complicate such a model of superconducting strings by
requiring a proliferation of scalar and spinor fields in order to obtain the
desired types of interactions. (It should be pointed out, however, that the
results obtained here apply to a specific Abelian gauge model, which is not
the only type of model that can describe superconducting strings with long
range gauge fields. For example, certain types of realistic grand unified
theories can also give rise to superconducting strings\cite{witten}$,$\cite
{everett}, where non-Abelian gauge field potentials are involved. It could
be of interest to investigate to what extent supersymmetry constrains more
realistic models.)

Next, the supersymmetric version of the $U(1)\times U(1)^{\prime }$-Higgs
model with the five chiral superfields $\Sigma _{+}$, $\Sigma _{-}$, $\Phi
_{+}$, $\Phi _{-}$, and ${\cal Z}$ has been investigated, and a
superpotential $W$ has been constructed which gives rise to a scalar
potential $V$ describing interactions among the bosonic components $\sigma
_{+}$, $\sigma _{-}$, $\phi _{+}$, $\phi _{-}$, and $Z$ of the chiral
superfields. In the bosonic sector of the theory, where the fermion fields
vanish, it is seen that when the bosonic fields $\phi _{\pm }$ and $Z$
vanish identically, and when a simplifying ansatz is used where $\sigma _{-}=%
\bar{\sigma}_{+}$ and $\phi _{-}=\bar{\phi}_{+}$, the model admits an
ordinary Nielsen-Olesen Abelian gauge string surrounded by a supersymmetric
vacuum. The two fields $\sigma _{+}$ and $\sigma _{-}$ thus conspire to form
the gauge string entrapping a unit of flux of the $B_\mu $ field. The
stability of the fields $\phi _{\pm }$ and $Z$ in the gauge string
background has then been examined in a first approximation by disregarding
the possible back reactions of the fields $\phi _{\pm }$ and $Z$ upon one
another. It has been determined that, for an appropriate parameter range,
the system must stabilize through the formation of a bosonic condensate of
the $\phi _{\pm }$ fields, while the $Z$ field, rather than forming a
condensate, can be associated with string-$Z$-particle bound states
localized within or near the string core. One also expects string-$Z$%
-particle scattering states to exist as well, and transitions between states
may occur due to $Z$ interactions with $\sigma _{\pm }$ and $\phi _{\pm }$
excitations. A nonvanishing current density $J_{A\mu }$, constructed from
the $\phi _{\pm }$ and $A_\mu $ fields, can then endow the string with
charge and/or current which generate the long range electromagnetic gauge
fields. Thus, although the two types of solutions are generally distinct,
the bosonic superconducting string solution obtained from the supersymmetric
version of the $U(1)\times U(1)^{\prime }$-Higgs model is seen to exhibit
many of the essential qualitative features of the superconducting string
solution emerging from the nonsupersymmetric version of the model.

\appendix

\section{Conventions}

Some of the notations and conventions are briefly listed here. A metric $%
g_{\mu \nu }$ is used with signature $(+,-,-,-)$. Aside from the metric, the
notation, conventions, and gamma matrices used conform to those of ref.\cite
{sbook}. The gamma matrices can be written in the form
\begin{equation}
\gamma ^\mu =i\left(
\begin{array}{cc}
0 & \sigma ^\mu \\
\bar{\sigma}^\mu & 0
\end{array}
\right)  \label{a1}
\end{equation}

\noindent with
\begin{equation}  \label{a2}
\sigma ^\mu =(1,{\bf \vec \sigma })\,\,,\,\,\,\,\,\,\,\,\,\,\bar \sigma ^\mu
=(1,-{\bf \vec \sigma })\,\,,
\end{equation}

\noindent where ${\bf \vec \sigma }$ represents the Pauli matrices. Then
\begin{equation}  \label{a3}
\gamma ^0=i\left(
\begin{array}{cc}
0 & 1 \\
1 & 0
\end{array}
\right) ,\,\,\,\,\,\,\,\,\,\,\gamma ^k=i\left(
\begin{array}{cc}
0 & \sigma _k \\
-\sigma _k & 0
\end{array}
\right) ,\,\,\,\,\,k=1,2,3,
\end{equation}

\noindent and $\gamma _5$ is given by
\begin{equation}  \label{a4}
\gamma _5=\gamma ^0\gamma ^1\gamma ^2\gamma ^3=i\left(
\begin{array}{cc}
1 & 0 \\
0 & -1
\end{array}
\right) .
\end{equation}

\noindent The gamma matrices have the properties
\begin{equation}  \label{a5}
\{\gamma ^\mu ,\gamma ^\nu \}=-2g^{\mu \nu },\,\,\,\,\{\gamma ^\mu ,\gamma
_5\}=0,\,\,\,\,\gamma _5^{\dagger }=-\gamma _5,\,\,\,\,(\gamma _5)^2=-1.
\end{equation}

\noindent A Majorana 4-spinor $\Psi $ is expressed in terms of the Weyl
2-spinors $\psi $ and $\bar \psi $ by $\Psi =\left(
\begin{array}{c}
\psi _\alpha \\
\bar \psi ^{\dot \alpha }
\end{array}
\right) $ and we use the summation conventions for Weyl spinors [with $\bar
\psi ^{\dot \alpha }=(\psi ^\alpha )^{*}$]
\begin{equation}  \label{a6}
\xi \psi \equiv \xi ^\alpha \psi _\alpha ,\,\,\,\,\bar \xi \bar \psi \equiv
\bar \xi _{\dot \alpha }\bar \psi ^{\dot \alpha },\,\,\,\,\alpha
=1,2,\,\,\,\,\dot \alpha =1,2,
\end{equation}

\noindent with $\varepsilon $ metric tensors (for raising and lowering Weyl
spinor indices)
\begin{equation}
(\varepsilon ^{\alpha \beta })=(\varepsilon ^{\dot{\alpha}\dot{\beta}%
})=i\sigma _2,\,\,(\varepsilon _{\alpha \beta })=(\varepsilon _{\dot{\alpha}%
\dot{\beta}})=-i\sigma _2,\,\,\,\,\varepsilon ^{12}=1=\varepsilon ^{\dot{1}%
\dot{2}}.  \label{a7}
\end{equation}

\newpage\

\end{document}